\documentclass{aa}

\usepackage{amssymb,amsmath,mathrsfs}
\usepackage{multirow}
\usepackage{graphicx}
\usepackage{float}
\usepackage{color}
\usepackage{ulem}
\usepackage{natbib}
\usepackage{wasysym}
\usepackage{txfonts}
\usepackage[latin1]{inputenc}

\def\etal{\textit{et al. }}

\begin{document}

\title{The rotation of Mimas}
\author{B.~Noyelles\thanks{F.R.S.-FNRS post-doctoral research fellow}\inst{1,2} \and \"O.~Karatekin\inst{3} \and N.~Rambaux\inst{4,2}}

\institute{FUNDP-University of Namur - Department of Mathematics \& NAmur Center for Complex SYStems (NAXYS) - Rempart de la Vierge 8 - B-5000 Namur - Belgium
\and IMCCE, Paris Observatory, UPMC, Univ. Lille 1, CNRS UMR 8028 - 77 avenue Denfert Rochereau - F-75014 Paris - France
\and Royal Observatory of Belgium - Ringlaan 3 - B-1180 Brussels - Belgium
\and Universit\'e Pierre et Marie Curie Paris 6 - 4 place Jussieu - F-75005 Paris - France}

\offprints{B.~Noyelles, e-mail: benoit.noyelles@fundp.ac.be}

\abstract{The Cassini mission in the Saturnian system is an outstanding opportunity to improve our knowledge of the satellites of 
Saturn. The data obtained thanks to this mission must be confronted to theoretical models.}{This paper aims at modeling the 
rotation of Mimas, with respect to its possible internal structure.}{For that, we first build different interior models, in 
considering Mimas as composed of 2 rigid layers with different porosity. Then we simulate the rotational behavior of these 
models in a 3-degree of freedom numerical code, in considering complete ephemerides of a Mimas whose rotation is disturbed 
by Saturn. We also estimate the deviation of its longitudinal orientation due to tides.}{We expect a signature of the internal 
structure up to $0.53^{\circ}$ in the longitudinal librations and an obliquity between 2 and 3 arcmin, the exact values depending 
on the interior.}{The longitudinal librations should be detectable, inverting them to get clues on the internal structure of 
Mimas is challenging.}


\date{Received / Accepted}

\maketitle

\section{Introduction}

The Cassini spacecraft gives us the unique opportunity to have accurate set of geodetic data for icy satellites of Saturn as 
for example, the shape, the gravitational field, the rotational data \citep{t10}. The flybys of Mimas have provided high 
resolution images of the surface in the finest detail yet seen \citep{rwhkmswdnhp09}. Cassini spacecraft has detected temperature 
inhomogeneities \citep{hssjphvs11}, usually attributed to exogenic process. The theoretical model of Mimas rotational state can be used 
to interpret the Cassini data and to better understand its interior and evolution.

Like for our Moon, Mimas is in synchronous rotation and shows almost the same face towards Saturn. Moreover, it is considered to have 
a large librational amplitude \citep{cb03}. The rotational state of a synchronous body depends on the distribution of mass of 
the body, and therefore it is a signature of its internal structure. Here, we propose to model the rotation of Mimas considering it 
as a rigid body. A highly rigid interior of Mimas  for most of its history is consistent with its un-relaxed shape 
\citep{tbhsvptmdgrjj07,t10}.

Since the distant spacecraft flybys of Mimas do not allow the determination of the GM nor the gravity harmonics, the mass of Mimas
is determined from an analysis of its orbital resonances with Tethys and Methone \citep{jabcijmpors06}. Moreover, its internal
structure remains uncertain. The mean density of $1.15$ $g$ $cm^{-3}$ suggests that Mimas is made of homogenous mixture of ice and 
rocks. The observed shape of Mimas by Voyager has been interpreted as an indication of interior mass concentration which can be 
either due to internal differentiation \citep{dt88} or radially variable porosity \citep{e90}. However, Cassini observations 
showed that Mimas' shape, although a triaxial ellipsoid, is departed slightly from hydrostatic shape and therefore interpreting the 
interior configuration from the shape is limited \citep{tbhsvptmdgrjj07,t10}. In the present study, we consider Mimas to be composed 
of two rigid layers. We consider both hydrostatic and non-hydrostatic interior models. The interior models considering compaction of
ice-silica particle mixtures \citep{ya09} are expected to yield realistic principal moment of inertia  $A<B<C$ \citep{e90}. Since 
Mimas orbits close to its parent planet, the present-day diurnal tidal stresses can be important, and we took the tidal effects 
into account as well.


The paper is structured as follows: we first model the internal structure of Mimas, in considering two different assumptions: that 
Mimas is in hydrostatic equilibrium, and that its ellipsoid of gravity is proportional to its ellipsoid of shape. From these 
two assumptions we derive 23 models of Mimas. Then we perform numerical integrations of the rotation of these "Mimasses" in a 
full 3-degree of freedom conservative models. Finally, we check the influence of the tides on the equilibrium orientation of 
Mimas' long axis.

\section{Internal structure}\label{sec:is}



\par Interior structure models of planets and natural satellites are in general non-unique due to the presence of fewer 
constraints than unknowns. For Mimas we have only two constraints: the mean radius $R$ and the mean density $\rho$, or the 
Mass $m$ (Tab.\ref{tab:propmim}). In this study where we deal with the rotation, the moment of inertia differences are 
the main point of interest.

\begin{table*}[ht]
\centering
\caption{Physical and dynamical properties of Mimas, used in the calculations. The mean density $\rho$ has been calculated from the radius of \citet{tbhsvptmdgrjj07} and the mass of \citet{jabcijmpors06}. A recent paper by \citet{t10} slightly shifts the mean radius to $198.3$, the change has negligible effects. However, we use this last reference for the triaxial shape of Mimas, because the rotation, especially the longitudinal motion, is sensitive to the differences between these axes.\label{tab:propmim}}
\begin{tabular}{l|ll}
\hline\hline
Parameter & Value & Source \\
\hline
Mean motion $n$ & $2435.14429644$ rad/y & TASS1.6 \citep{vd95} \\
Mean radius $R$ & $198.2$ km & \citep{tbhsvptmdgrjj07} \\
Density $\rho$ & $1150.03$ $kg.m^{-3}$ & \citep{tbhsvptmdgrjj07} \\
Mass $m$ & $3.7495\times10^{19}$ $kg$ & \citep{jabcijmpors06} \\
Saturn-facing radius $a$ & $207.8$ km & \citep{t10} \\
Orbit facing radius $b$ & $196.7$ km & \citep{t10} \\
Polar radius $c$ & $190.6$ km & \citep{t10} \\
\hline
\end{tabular}
\end{table*}

\par Because we have only few constraints, we prefer to have as simple interior structure models as possible. We assume a 
two-layer interior structure model with a rocky core and icy mantle. The models with variable porosity supported by recent 
compaction experiments \citep{ya09}, yield realistic moments of inertia \citep{e90}. The mean density of Mimas is close to 
the density of water ice and the interior is consistent with an icy mantle and small rocky core, alternatively Mimas can be 
homogenous with a variable or constant porosity. 

\par With Mimas' figure departed from hydrostatic shape \citep{tbhsvptmdgrjj07,t10} and without the knowledge of $C_{22}$ and $J_2$, 
we do not know the moment of inertia differences. As it is the case for the Moon (see e.g. \citet{lp80}), the internal mass 
distribution could be a a fossil shape which dates back from an earlier orbital position where the tidal heating was important 
with higher orbital eccentricity and obliquity or a frozen shape following a large impact. Mimas shows a heavily cratered 
surface without signs of geological activity for billions of years. Its large free eccentricity gives another reason for low 
internal activity. This anomalously large eccentricity of Mimas can be explained by passage through several past resonances 
\citep{mw08}. Determination of the gravity coefficients $C_{22}$ and $J_2=-C_{20}$ are necessary to conclude on the hydrostatic 
equilibrium because the figure of Mimas may not represent the real flattening or the internal mass determination as it is the 
case for Titan \citep{zshlkl09}. As suggested by \citet{jcm06}, large impact craters and heating of Mimas' hemispheres by Saturn 
at different amounts may be potential sources of large non-hydrostatic anomalies that could impede accurate interpretation of 
the shape data

\subsection{Hydrostatic approximation}

\par Since we do not know the gravity coefficients $C_{22}$ and $J_2$, we will use a simple approach. For a two layer interior model 
the core radius $R_c$ can be determined if the densities of the rocky core $\rho_c$ and the icy mantle $\rho_s$ are known, i.e.
 
\begin{equation}
\label{equ:rc}
R_c=R\left(\frac{\rho-\rho_s}{\rho_c-\rho_s}\right)^{1/3}.
\end{equation}
The moment of Inertia factor ($MOI=I_p/(MR^2)$) is given as:

\begin{equation}
MOI=\frac{2}{5}\left(\frac{\left(\rho-\rho_s\right)^{5/3}}{\rho\left(\rho_c-\rho_s\right)^{2/3}}+\frac{\rho_s}{\rho}\right).
\end{equation}

\begin{figure}
\includegraphics[width=6cm,height=4.5cm]{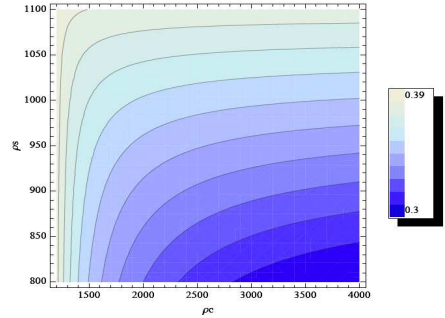}
\caption[The Moment of Inertia factor]{Variations of the Moment of Inertia factor (MOI) with icy shell and rocky core densities.\label{fig:moi}}
\end{figure}

The Fig.\ref{fig:moi} shows that in the absence of additional constraints, plausible density values of the core and icy shell 
yield $0.3 < MOI < 0.4$. The range of MOI is used to estimate the moment of inertia of tri-axial Mimas, as shown below.

\par For a satellite in hydrostatic equilibrium, $MOI$ is related to the fluid Love number $k_f$ which describes the reaction of the 
satellite to a perturbing potential after all viscous stresses have relaxed \citep{mm60,ha78}:

\begin{equation}
\label{eq:moi}
MOI=\frac{2}{3}\left[1-\frac{2}{5}\sqrt{\frac{4-k_f}{1+k_f}}\right],
\end{equation}
and the gravity coefficients $C_{22}$ and $J_2$ are determined from \citep{rbga97}:

\begin{eqnarray}
C_{22} & = & \frac{k_f}{4}q_r+O\left(q_r^2\right), \label{eq:C22} \\
J_2 & = & \frac{5k_f}{6}q_r+O\left(q_r^2\right), \label{eq:J2}
\end{eqnarray}
where $q_r=\Omega^2R^3/(Gm)$, $\Omega$ being the spin velocity of Mimas, equal to its mean motion $n$ since Mimas is in synchronous rotation. With the numerical values given in Tab.\ref{tab:propmim}, we have $q_r=0.01854$.

\par The differences between the three principal moments of inertia $A<B<C$ are determined from the definitions of $C_{22}$ and $J_2$, i.e.:

\begin{eqnarray}
B-A & = & 4C_{22}MR^2, \nonumber \\
C-A & = & (J_2+2C_{22})MR^2, \nonumber \\
C-B & = & (J_2-2C_{22})MR^2. \nonumber
\end{eqnarray}
\par The relationship between the mean moment of inertia $I=\frac{A+B+C}{3}$ and the polar moment of inertia $C$ is:

\begin{equation}
C=I+\frac{2}{3}J_2MR^2.
\end{equation}
We can then calculate all the three moments of inertia $A$, $B$ and $C$.

\subsection{Nonhydrostatic shape}

\par We here use the observed shape ($a=207.8$ km, $b=196.7$ km, $c=190.6$ km, \citet{t10}) to calculate the moments of inertia of Mimas, in assuming that the shape of the core is proportional to the one of Mimas, i.e. we assume

\begin{equation}
\label{equ:propor}
\frac{a_c}{a}=\frac{b_c}{b}=\frac{c_c}{c}=\frac{R_c}{R},
\end{equation}
where $a_c$, $b_c$ and $c_c$ are the dimensions of the core, and $R_c$ its mean radius (Eq.\ref{equ:rc}).

\par A quadrature over the volume of respectively the core and the shell gives

\begin{eqnarray}
C_c & = & \iiint_{core}\rho_c\left(x^2+y^2\right)dx\,dy\,dz \nonumber \\
 & = & \frac{4}{15}\pi a_cb_cc_c\left(a_c^2+b_c^2\right)\rho_c \nonumber \\
 & = & \frac{4}{15}\pi abc\left(a^2+b^2\right)\rho_c\left(\frac{R_c}{R}\right)^5, \label{eq:cc} \\
C_s & = & \iiint_{Mimas}\rho_s\left(x^2+y^2\right)dx\,dy\,dz \nonumber \\
& & -\iiint_{core}\rho_s\left(x^2+y^2\right)dx\,dy\,dz \nonumber \\
 & = & \frac{4}{15}\pi abc\left(a^2+b^2\right)\rho_c\left[1-\left(\frac{R_c}{R}\right)^5\right].\label{eq:cs}
\end{eqnarray}
We then get $C=C_c+C_s$. The other moments of inertia $A$ and $B$ being obtained similarly, we have

\begin{eqnarray}
A & = & \frac{4}{15}\pi abc\left(b^2+c^2\right)\left[\left(\rho_c-\rho_s\right)\left(\frac{R_c}{R}\right)^5+\rho_s\right], \label{eq:ga} \\
B & = & \frac{4}{15}\pi abc\left(a^2+c^2\right)\left[\left(\rho_c-\rho_s\right)\left(\frac{R_c}{R}\right)^5+\rho_s\right], \label{eq:gb} \\
C & = & \frac{4}{15}\pi abc\left(a^2+b^2\right)\left[\left(\rho_c-\rho_s\right)\left(\frac{R_c}{R}\right)^5+\rho_s\right]. \label{eq:gc}
\end{eqnarray}
We finally see that the ratio of the moments of inertia 
$A/C=\left(b^2+c^2\right)/\left(a^2+b^2\right)$ and $B/C=\left(a^2+c^2\right)/\left(a^2+b^2\right)$ are independent on the mean 
radius $R_c$ and density $\rho_c$ of the core, so every model of the internal structure of Mimas based on its observed shape 
(in neglecting the uncertainties on the radii $a$, $b$ and $c$) will present the same rotational response.

\par The interior models considered in the present study are gathered in Tab.\ref{tab:lescas}.

\begin{table*}[ht]
\centering
\caption{The interior models considered in the present study. The first 22 cases have been elaborated in considering Mimas as in 
hydrostatic equilibrium, while the $23^{rd}$ is based on the observed shape. We give only one possibility for the shape model 
because the ratios of the moments of inertia $A/C$ and $B/C$ remain constant, so the rotational response of Mimas is the same for any interior model based on the shape.\label{tab:lescas}}
\begin{tabular}{l|lllllll}
\hline\hline
N & $\rho_c$ & $\rho_s$ & $k_f$ & $MOI$ & $J_2$ $(10^{-2})$ & $C_{22}$ $(10^{-3})$ & $C/(mR^2)$ \\
\hline
 1 & $1200$ &  $800$ & $1.40473$ & $0.389636$ & $2.17051$ & $6.51152$ & $0.404106$ \\
 2 & $1500$ &  $800$ & $1.11293$ & $0.354953$ & $1.71963$ & $5.15889$ & $0.366418$ \\
 3 & $2000$ &  $800$ & $0.94032$ & $0.331801$ & $1.45293$ & $4.35878$ & $0.341487$ \\
 4 & $2500$ &  $800$ & $0.86349$ & $0.320705$ & $1.33422$ & $4.00267$ & $0.329600$ \\
 5 & $3000$ &  $800$ & $0.81885$ & $0.314001$ & $1.26524$ & $3.79571$ & $0.322436$ \\
 6 & $3500$ &  $800$ & $0.78921$ & $0.309439$ & $1.21944$ & $3.65831$ & $0.317569$ \\
 7 & $4000$ &  $800$ & $0.76788$ & $0.306100$ & $1.18649$ & $3.55946$ & $0.314010$ \\
\hline
 8 & $1200$ & $1000$ & $1.41613$ & $0.390899$ & $2.18812$ & $6.56437$ & $0.405486$ \\
 9 & $1500$ & $1000$ & $1.24455$ & $0.371206$ & $1.92301$ & $5.76902$ & $0.384026$ \\
10 & $2000$ & $1000$ & $1.17336$ & $0.362551$ & $1.81300$ & $5.43901$ & $0.374638$ \\
11 & $2500$ & $1000$ & $1.14536$ & $0.359061$ & $1.76975$ & $5.30925$ & $0.370860$ \\
12 & $3000$ & $1000$ & $1.12980$ & $0.357099$ & $1.74570$ & $5.23711$ & $0.368737$ \\
13 & $3500$ & $1000$ & $1.11969$ & $0.355816$ & $1.73009$ & $5.19026$ & $0.367350$ \\
14 & $4000$ & $1000$ & $1.11251$ & $0.354901$ & $1.71899$ & $5.15698$ & $0.366361$ \\
\hline
15 & $1200$ & $1100$ & $1.44040$ & $0.393565$ & $2.22563$ & $6.67688$ & $0.408403$ \\
16 & $1500$ & $1100$ & $1.38066$ & $0.386951$ & $2.13331$ & $6.39993$ & $0.401173$ \\
17 & $2000$ & $1100$ & $1.36451$ & $0.385134$ & $2.10836$ & $6.32508$ & $0.399189$ \\
18 & $2500$ & $1100$ & $1.35879$ & $0.384487$ & $2.09953$ & $6.29858$ & $0.398484$ \\
19 & $3000$ & $1100$ & $1.35572$ & $0.384139$ & $2.09478$ & $6.28435$ & $0.398105$ \\
20 & $3500$ & $1100$ & $1.35376$ & $0.383917$ & $2.09176$ & $6.27527$ & $0.397862$ \\
21 & $4000$ & $1100$ & $1.35239$ & $0.383761$ & $2.08963$ & $6.26890$ & $0.397692$ \\
\hline
22 & $1150.03$ & $1150.03$ & $1.5$ & $0.400000$ & $2.30951$ & $6.92854$ & $0.415397$ \\
\hline
23 & $1200$ &  $800$ & $1.40473$ & $0.389636$ & $2.28639$ & $5.57013$ & $0.406273$ \\
\hline
 \end{tabular}
\end{table*}

\section{Computing the rotation of Mimas}

In this Section, Mimas is assumed to be a two-layers rigid body and the tidal contributions will be investigated in the 
Section~\ref{sec:tidal}. Its rotation is highly constrained by the gravitational perturbation of Saturn, and so depends on the variations of the distance Mimas-Saturn. That is the reason why we must understand the orbital dynamics of Mimas before investigating its rotation.

\subsection{The orbital dynamics of Mimas}

Mimas is the smallest of the main Saturnian satellites, and also the closest to its parent planet and the rings. Discovered by 
Herschel in 1789, it is known since \citet{s91} to be in 2:1 mean-motion resonance with Tethys. More precisely, these two bodies 
are locked in an inclination-type resonance whose argument is $2\lambda_1-4\lambda_3+\ascnode_1+\ascnode_3$, the subscript $1$ 
standing for the satellite S-1 Mimas, $3$ for S-3 Tethys, $\lambda_i$ being the mean longitudes, and $\ascnode_i$ the 
longitudes of the ascending nodes. This resonance tends to raise the inclinations of the satellites to $\approx1.5^{\circ}$ for 
Mimas and $\approx1^{\circ}$ for Tethys \citep{a69}, and stimulates librations of the resonant argument around $0$ with an 
amplitude of $\approx95^{\circ}$ and a period of $\approx70$ years. The trapping of the system into this resonance can be 
explained in considering a non-null eccentricity for Tethys that induces secondary resonances that strongly enhances the capture 
probability \citep{cv99a,cv99b}.

It is convenient to work on a Fourier-type representation of the orbital motion of Mimas that allows to identify every proper mode of the motion. The basic idea is that the variables describing the orbital motion of Mimas can be represented as quasi-periodic series (and a slope for precessing angles like the ascending node, the pericenter and the mean longitude), i.e. infinite but converging sums of trigonometric series. The arguments of these series can be expressed as integer combination of a few proper modes of constant frequencies. The existence of these modes comes both from the KAM \citep{a63,m62} and the Nekhoroshev theories \citep{n77,n79}. The KAM theory states that for a quasi-integrable Hamiltonian system (i.e. like $\mathcal{H}=\mathcal{H}_0+\epsilon\mathcal{H}_1$ where $\mathcal{H}_0$ is an integrable Hamiltonian and $\epsilon\mathcal{H}_1$ a small perturbation) verifying classical assumptions, the motion can be considered to be on invariant tori (i.e. with constant amplitudes and angles depending linearly on time) in action-angle coordinates. For a bigger perturbation the Nekhoroshev theory says that the invariant tori survive over a timescale that is exponentially long with respect to the invert of the amplitude of the perturbation $\epsilon$, provided that the Hamiltonian of the system presents a property of steepness, that is an extension of the convexity.

Such a representation is given by TASS1.6 ephemerides \citep{vd95} where the orbital motion of Mimas can be described using the 
5 proper modes $\lambda$, $\omega$, $\phi$, $\zeta$ and $\Phi$. $\lambda$ is the linear part of Mimas' mean longitude, $\omega$ is 
the main oscillation mode of the librations of the resonant argument $2\lambda_1-4\lambda_3+\ascnode_1+\ascnode_3$, $\zeta$ 
(called $\rho_1$ in \citep{vd95}) is the mean slope of $\lambda_1-2\lambda_3$, and $\phi-\zeta$ and $\Phi-\zeta$ are the mean slopes 
of respectively the longitudes of the pericenter of Mimas and its ascending node. The values of the frequency associated are 
gathered in Tab.\ref{tab:propmod}.

\begin{table}[ht]
\centering
\caption{The proper frequencies of Mimas' orbital motion (from TASS1.6 \citep{vd95}).\label{tab:propmod}}
\begin{tabular}{l|lll}
\hline\hline
 & Frequency (rad/y) & Period (d) & Period (y) \\
\hline
$\lambda$ & $2435.14429644$ & $0.942421949$ & $2.580211\times10^{-3}$ \\
$\omega$ &     $0.08904538$ & $25772.62777$ & $70.561609$ \\
$\phi$ &      $10.19765304$ & $225.0452555$ & $0.616140$ \\
$\zeta$ &       $3.81643833$ & $601.3285779$ & $1.646348$ \\
$\Phi$ &      $-2.55544336$ & $898.0568575$ & $2.458746$ \\
\hline
\end{tabular}
\end{table}

\subsection{Rotational model}

\par As for most of the natural satellites of the Solar System, Mimas is expected to follow the 3 Cassini Laws, originally described 
for the Moon \citep{c93,c66}, i.e.:

\begin{enumerate}

\item The Moon rotates uniformly about its polar axis with a rotational period equal to the mean sidereal period of its orbit about the Earth.

\item The inclination of the Moon's equator to the ecliptic is a constant angle (approximately $1.5^{\circ}$).

\item The ascending node of the lunar orbit on the ecliptic coincides with the descending node of the lunar equator on the ecliptic. This law could also be expressed as: the spin axis and the normals to the ecliptic and orbit plane remain coplanar.

\end{enumerate}
In the case of natural satellites, they can be rephrased this way: the rotation of the satellite is synchronous, its angular momentum has a nearly constant inclination on an inertial reference plane, and is located in the plane defined by the normal to the orbital plane and to the Laplace Plane. The Laplace Plane is the plane normal to the rotation axis of the orbital frame, i.e. it is defined with respect to the orbital precessional motion. It has the property to minimize the variations of the orbital inclinations. For satellites orbiting close to their planet as it is the case here, the equatorial plane of Saturn is so close to the Laplace Plane that it can be used for describing the rotational dynamics.


\par Our rotational model is similar to the one already used in e.g. \citep{nlv08,n10} for studying the rigid rotation of the Saturnian satellites Titan, Janus and Epimetheus.

We consider Mimas as a rigid triaxial body whose matrices of inertia reads

\begin{equation}
I=\left(\begin{array}{ccc}
A & 0 & 0 \\
0 & B & 0 \\
0 & 0 & C
\end{array}\right)
\label{equ:inertie}
\end{equation}
with $A \leq B \leq C$.

\par The dynamical model is a 3-degree of freedom one. We will use the Andoyer variables which requires a decomposition with 3 references frames :
\begin{enumerate}

\item An inertial reference frame $(\vec{e_1},\vec{e_2},\vec{e_3})$. We used the one in which the orbital ephemerides are given, i.e. mean Saturnian equator and mean equinox for J2000.0 epoch.

\item A frame $(\vec{n_1},\vec{n_2},\vec{n_3})$ bound to the angular momentum of Mimas.

\item A frame $(\vec{f_1},\vec{f_2},\vec{f_3})$ rigidly linked to Mimas. 

\end{enumerate}

\begin{figure}
\includegraphics[width=8.5cm,height=4.5cm]{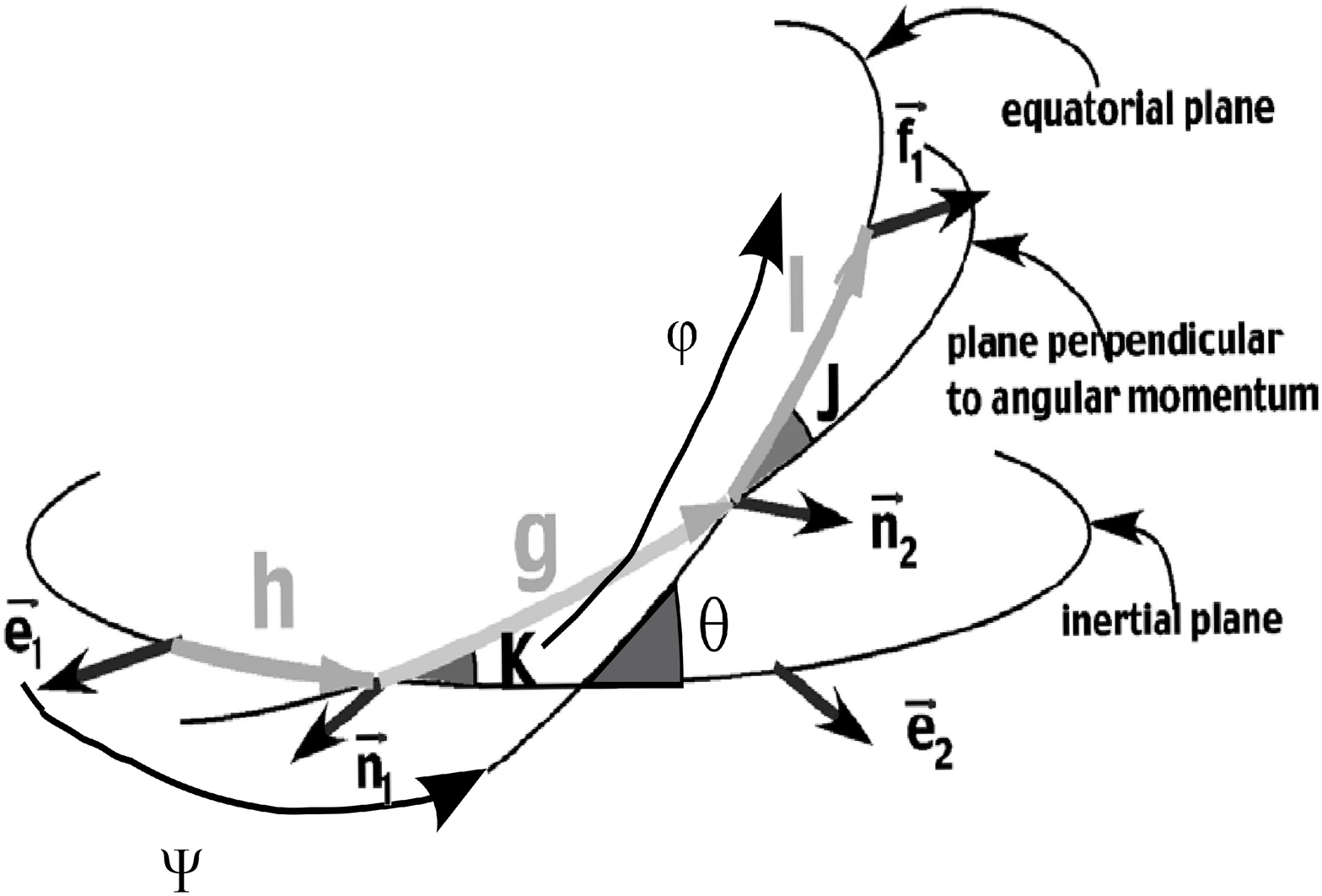}
\caption[The Andoyer variables]{The Andoyer variables (partially reproduced from \citep{h05io}).}
\label{fig01}
\end{figure}

\par We first use Andoyer's variables \citep{a26,d67}, which are based on two linked sets of Euler's angles. The first set $(h,K,g)$ locates the position of the angular momentum in the first frame $(\vec{e_1},\vec{e_2},\vec{e_3})$, while the second one, $(g,J,l)$, locates the body frame $(\vec{f_1},\vec{f_2},\vec{f_3})$ in the second frame tied to the angular momentum (see Fig. \ref{fig01}).

\par The canonical set of Andoyer's variables consists of the three angular variables $l,g,h$ and their conjugated momenta $L,G,H$ defined by the norm $G$ of the angular momentum and two of its projections: 

\begin{equation}
\begin{array}{lll}
l, & \hspace{3cm} & L=G\cos J, \\
g, & \hspace{3cm} & G, \\
h, & \hspace{3cm} & H=G\cos K.
\end{array}
\end{equation}

\par Unfortunately, these variables present two singularities: when $J=0$ (i.e., the angular momentum is colinear to $\vec{f_3}$), $l$ and $g$ are undefined, and when $K=0$ (i.e., when Mimas' principal axis of inertia is perpendicular to its orbital plane), $h$ and $g$ are undefined. That is the reason why we shall use the modified Andoyer's variables:

\begin{equation}
\begin{array}{lll}
p=l+g+h, & \hspace{2cm} & P=\frac{G}{nC}, \\
\vspace{0.3cm}
r=-h, & \hspace{2cm} & \mathcal{R}=\frac{G-H}{nC}=P(1-\cos K), \\
\vspace{0.3cm}
& \hspace{2cm} & =2P\sin^2\frac{K}{2}, \\
\vspace{0.3cm}
\xi_q=\sqrt{\frac{2Q}{nC}}\sin q, & \hspace{2cm} & \eta_q=\sqrt{\frac{2Q}{nC}}\cos q, \label{equ:modified} \\
\end{array} \\
\end{equation}
where $n$ is the body's mean orbital motion , $q=-l$, and $Q=G-L=G(1-\cos J)=2G\sin^2\frac{J}{2}$. With these new variables, the singularity on $l$ has been dropped. Using these variables has a great mathematical interest, because they are canonical, so they simplify an analytical study of the system, as was done in the previous works mentioned above. Our study here is essentially numerical, but we keep these variables, in order to be consistent with previous studies. We later derive other output variables, that are more relevant from a physical point of view.

\par In these variables, the kinetic energy $T=\frac{1}{2}\vec{\omega}\cdot\vec{G}$ of the system reads:

\begin{eqnarray}
\label{equ:kinrj}
T & = & \frac{nP^2}{2}+\frac{n}{8}\big[4P-\xi_q^2-\eta_q^2\big] \nonumber \\ 
& \times & \Big[\frac{\gamma_1+\gamma_2}{1-\gamma_1-\gamma_2}\xi_q^2+\frac{\gamma_1-\gamma_2}{1-\gamma_1+\gamma_2}\eta_q^2\Big]
\end{eqnarray}
with 

\begin{equation}
\gamma_1=\frac{2C-A-B}{2C}=J_2\frac{MR^2}{C}
\label{equ:gama1}
\end{equation}
and

\begin{equation}
\gamma_2=\frac{B-A}{2C}=2C_{22}\frac{MR^2}{C}.
\label{equ:gama2}
\end{equation}
In these last 3 formulae, $\vec{\omega}$ is the instantaneous vector of rotation, $M$ is the mass of Mimas, $R$ its mean radius, and $J_2$ and $C_{22}$ the two classical normalized gravitational coefficients related respectively to the oblateness and equatorial ellipticity of the considered body.

\par The gravitational disturbing potential due to an oblate perturber $p$ reads \citep{h05eu2}:

\begin{equation}
V_p=V_{p1}+V_{p2}
\label{equ:Vpert}
\end{equation}
with

\begin{equation}
V_{p1}=-\frac{3}{2}C\frac{\mathcal{G}M_p}{d_p^3}\big[\gamma_1(x_p^2+y_p^2)+\gamma_2(x_p^2-y_p^2)\big]
\label{equ:Vpert1}
\end{equation}
and

\begin{eqnarray}
V_{p2} & = & -\frac{15}{4}CJ_{2p}\frac{\mathcal{G}M_p}{d_p^3}\Big(\frac{R_p}{d_p}\Big)^2 \nonumber \\
& \times & \big[\gamma_1(x_p^2+y_p^2)+\gamma_2(x_p^2-y_p^2)\big],
\label{equ:Vpert2}
\end{eqnarray}
where $\mathcal{G}$ is the gravitational constant, $M_p$ the mass of the perturber, $J_{2p}$ its $J_2$, $R_p$ its mean radius, $d_p$ the distance between the perturber's and Mimas' centers of mass, and $x_p$ and $y_p$ the two first components of the unit vector pointing to the center of mass of the perturber, from the center of mass of the body, in the reference frame $(\vec{f_1},\vec{f_2},\vec{f_3})$. $V_{p1}$ expresses the perturbation due to a pointmass perturber, while $V_{p2}$ represents the perturbation due to its $J_2$, assuming that the body is in the equatorial plane of the perturber. As shown in \citep{h05eu2}, it is a good approximation if the sine of the angle between Saturn's equatorial plane and the orbit is small. In the case of Mimas, this angle (i.e. Mimas' orbital inclination) is $\approx1.5^{\circ}\approx2.6\times10^{-2}$ rad, so we can consider that its sine is always smaller than $3\times10^{-2}$. This assertion also assumes that the obliquity of Mimas is very small, what will be checked in this study. 

\par Usually the orbital ephemerides give us the location of the perturber in the inertial frame, so we have to perform 5 rotations to convert the coordinates from the inertial frame to $(\vec{f_1},\vec{f_2},\vec{f_3})$. More precisely, if we name $(x_i,y_i,z_i)^T$ the unit vector locating the perturber's center of mass in the inertial frame, we have 

\begin{equation}
\left(\begin{array}{c}
x_p \\
y_p \\
z_p
\end{array}\right)
=R_3(-l)R_1(-J)R_3(-g)R_1(-K)R_3(-h)\left(\begin{array}{c}
x_i \\
y_i \\
z_i
\end{array}\right)
\label{equ:passage}
\end{equation}
with

\begin{equation}
R_3(\phi)=\left(\begin{array}{ccc}
\cos\phi & -\sin\phi & 0 \\
\sin\phi & \cos\phi & 0 \\
0 & 0 & 1
\end{array}\right)
\label{equ:r3}
\end{equation}
and

\begin{equation}
R_1(\phi)=\left(\begin{array}{ccc}
1 & 0 & 0 \\
0 & \cos\phi & -\sin\phi \\
0 & \sin\phi & \cos\phi 
\end{array}\right).
\label{equ:r1}
\end{equation}

\par Finally, the total Hamiltonian of the problem reads:

\begin{eqnarray}
H & = & \frac{nP^2}{2}+\frac{n}{8}\big[4P-\xi_q^2-\eta_q^2\big] \nonumber \\
 & & \times\Big[\frac{\gamma_1+\gamma_2}{1-\gamma_1-\gamma_2}\xi_q^2+\frac{\gamma_1-\gamma_2}{1-\gamma_1+\gamma_2}\eta_q^2\Big] \nonumber \\
 & & -\frac{3}{2n}\frac{\mathcal{G}M_{\saturn}}{d_{\saturn}^3}\Big(1+\frac{5}{2}J_{2{\saturn}}\Big(\frac{R_{\saturn}}{d_{\saturn}}\Big)^2\Big) \nonumber \\
 & & \times  \big[\gamma_1(x_{\saturn}^2+y_{\saturn}^2)+\gamma_2(x_{\saturn}^2-y_{\saturn}^2)\big],
\label{equ:Htotal}
\end{eqnarray}
where the index $\saturn$ stands for Saturn. We will use this Hamiltonian for a numerical study of the rotation. An analytical study can show that the Hamiltonian (\ref{equ:Htotal}) can be reduced to 

\begin{equation}
\mathcal{H}(u,v,w,U,V,W)=\omega_uU+\omega_vV+\omega_wW+\mathcal{P}(u,v,w,U,V,W)
\label{equ:quadra}
\end{equation}
where $\mathcal{P}$ represents a perturbation, and the three constants $\omega_u$, $\omega_v$ and $\omega_w$ are the periods of the free oscillations around the equilibrium defined by the Cassini Laws. This last Hamiltonian is obtained after several canonical transformations, the first one consisting in expressing the resonant arguments $\sigma=p-\lambda+\pi$ and $\rho=r+\ascnode$ respectively associated with the 1:1 spin-orbit resonance and with the orientation of the angular momentum, $\lambda$ and $\ascnode$ being the orbital variables defined above. The complete calculation is beyond the scope of this paper, the reader can find details in \citep{h05io,h05eu,nlv08}.

\subsection{A numerical study}

\par In order to integrate numerically the system, we first express the coordinates of the perturber $(x_{\saturn},y_{\saturn})$ with the numerical ephemerides and the rotations given in (Eq.\ref{equ:passage}), in the body frame $(\vec{f_1},\vec{f_2},\vec{f_3})$. As explained before, the ephemerides are given by the TASS1.6 ephemerides \citep{vd95}. This way, we get coordinates depending of the canonical variables. Then we derive the equations coming from the Hamiltonian (\ref{equ:Htotal}):

\begin{eqnarray}
\frac{dp}{dt} =  \frac{\partial H}{\partial P}, & &  \frac{dP}{dt} = -\frac{\partial H}{\partial p},  \nonumber \\
\frac{dr}{dt} = \frac{\partial H}{\partial R}, & & \frac{dR}{dt}=-\frac{\partial H}{\partial r}, \nonumber \\
\frac{d\xi_q}{dt} = \frac{\partial H}{\partial \eta_q}, & & \frac{d\eta_q}{dt}=-\frac{\partial H}{\partial \xi_q}. \label{equ:equhamil}
\end{eqnarray}

\par We integrated over 200 years using the Adams-Bashforth-Moulton 10th order predictor-corrector integrator. The solutions consist of two parts, the forced one, directly due to the perturbation, and the free one, that depends on the initial conditions. The initial conditions should be as close as possible to the exact equilibrium, that is assumed to be the Cassini State $1$ in 1:1 spin-orbit resonance, to have low amplitudes of the free librations. For that, we have used the iterative algorithm NAFFO \citep{ndc11} to remove the free librations from the initial conditions, after they have been identified by frequency analysis.

The frequency analysis algorithm we used is based on Laskar's original idea, named NAFF as Numerical Analysis of the Fundamental Frequencies (see for instance \citet{l93} for the method, and \citet{l05} for the convergence proofs). It aims at identifying the coefficients $a_k$ and $\omega_k$ of a complex signal $f(t)$ obtained numerically over a finite time span $[-T;T]$  and verifying

\begin{equation}
\label{equ:naff}
f(t) \approx \sum_{k=1}^na_k\exp(\imath\omega_kt),
\end{equation}
where $\omega_k$ are real frequencies and $a_k$ complex coefficients. If the signal $f(t)$ is real, its frequency spectrum is symmetric and the complex amplitudes associated with the frequencies $\omega_k$ and $-\omega_k$ are complex conjugates. The frequencies and amplitudes associated are found with an iterative scheme. To determine the first frequency $\omega_1$, one searches for the maximum of the amplitude of 

\begin{equation}
\label{equ:philas}
\phi(\omega)=<f(t),\exp(\imath\omega t)>,
\end{equation}
where the scalar product $<f(t),g(t)>$ is defined by

\begin{equation}
\label{equ:prodscal}
<f(t),g(t)>=\frac{1}{2T}\int_{-T}^T f(t)\overline{g(t)}\chi(t) dt,
\end{equation}
and where $\chi(t)$ is a weight function, i.e. a positive function with

\begin{equation}
\label{equ:poids}
\frac{1}{2T}\int_{-T}^T \chi(t) dt=1.
\end{equation}
Once the first periodic term $\exp(\imath\omega_1t)$ is found, its complex amplitude $a_1$ is obtained by orthogonal projection, and the process is started again on the remainder $f_1(t)=f(t)-a_1\exp(\imath\omega_1t)$. The algorithm stops when two detected frequencies are too close to each other, which alters their determinations, or when the number of detected terms reaches a maximum set by the user. This algorithm is very efficient, except when two frequencies are too close to each other. In that case, the algorithm is not confident in its accuracy and stops. When the difference between two frequencies is larger than twice the frequency associated with the length of the total time interval, the determination of each fundamental frequency is not perturbed by the other ones. Although the iterative method suggested by \citet{c98} allows to reduce this distance, some difficulties remain when the frequencies are too close to each other.

\subsection{Outputs}\label{sec:output}

\par In order to deliver theories of rotation that can be easily compared with observations, we chose to express our results in the following variables:

\begin{itemize}

\item Longitudinal librations,

\item Latitudinal librations,

\item Orbital obliquity $\epsilon$ (the orientation of the angular momentum of Mimas with respect to the normal to the instantaneous orbital plane),

\item Motion of the rotation axis about the pole axis.

\end{itemize}

\par There are at least two ways to define the longitudinal librations. We can for instance consider the librations about the exact synchronous rotation, i.e. $p-<n>t$, usually called \emph{physical librations}. In this case, we have used for $<n>$ the frequency associated with the proper mode $\lambda$, i.e. Mimas' mean longitude. Another way to consider the longitudinal librations is to work on the librations about the Mimas-Saturn direction. We will call these librations \emph{tidal librations} because they represent the misalignment of the tidal bulge of the satellite. The difference between these two librations is known as \emph{optical librations}, only due to the orbital motion of Mimas around Saturn. The reader can find graphical descriptions of these librations in \citet{md99}, Fig.5.16.

\par The latitudinal librations are the North-South librations of the large axis of the considered body in the saturnocentric reference frame that follows the orbital motion of the body. They are analogous to the tidal librations that are the East-West librations. In order to get the tidal longitudinal librations and the latitudinal librations, we first should express the unit vector $\vec{f_1}$ (i.e. the direction of Mimas' long axis) in the inertial frame $(\vec{e_1},\vec{e_2},\vec{e_3})$. From (Eq.\ref{equ:passage}) and the definitions of the Andoyer modified variables (Eq.\ref{equ:modified}), we get:

\begin{eqnarray}
\vec{f_1} & = & (\cos r (\cos(p+r-l)\cos l-\sin(p+r-l)\cos J\sin l) \nonumber \\
 & & +\sin r(\cos K(\sin(p+r-l)\cos l \nonumber \\
 & & +\cos (p+r-l)\cos J\sin l)-\sin K\sin J\sin l)) \vec{e_1} \nonumber \\
 & + & (-\sin r (\cos(p+r-l)\cos l-\sin(p+r-l)\cos J\sin l) \nonumber \\
 & & +\cos r(\cos K(\sin(p+r-l)\cos l \nonumber \\
 & & +\cos (p+r-l)\cos J\sin l)-\sin K\sin J\sin l)) \vec{e_2} \nonumber \\
 & + & (\sin K (\sin(p+r-l)\cos l+\cos(p+r-l)\cos J\sin l) \nonumber \\
 & & +\cos K\sin J\sin l) \vec{e_3}. \label{equ:f1}
\end{eqnarray}
The tidal longitudinal librations $\psi$ and the latitudinal ones $\eta$ are found this way:

\begin{equation}
\label{equ:psicross}
\psi=\vec{t}\cdot\vec{f_1}
\end{equation}
and

\begin{equation}
\label{equ:etacross}
\eta=\vec{n}\cdot\vec{f_1},
\end{equation}
where $\vec{n}$ is the unit vector normal to the orbit plane, and $\vec{t}$ the tangent to the trajectory. We get these last two vectors by:

\begin{equation}
\label{equ:vecnorm}
\vec{n}=\frac{\vec{x}\times\vec{v}}{||\vec{x}\times\vec{v}||}
\end{equation}
and

\begin{equation}
\label{equ:vectang}
\vec{t}=\frac{\vec{n}\times\vec{x}}{||\vec{n}\times\vec{x}||},
\end{equation}
where $\vec{x}$ is the position vector of the body, and $\vec{v}$ its velocity.

\par Finally, the motion of the rotation axis about the pole is derived from the wobble $J$, it is given by the two variables $Q_1$ and $Q_2$ defined as:

\begin{equation}
\label{equ:Q1}
Q_1=\sin J \sin l \bigg(1+\frac{J_2+2C_{22}}{C}\bigg)
\end{equation}
and

\begin{equation}
\label{equ:Q2}
Q_2=\sin J \cos l \bigg(1+\frac{J_2-2C_{22}}{C}\bigg),
\end{equation}
they are the first two components of the unit vector pointing at the instantaneous North Pole of Mimas' rotation axis, in the body frame of Mimas. These quantities are finally multiplied by the polar radius of the satellite, i.e. $190.6$ km \citep{t10} to get a deviation in meters.

\section{Results}

\par We here present the outputs of our numerical study of the rotation of Mimas. We first give the example of a non-hydrostatic
model of Mimas based on its observed shape, then we compare the results with the rotational response of the first 22 models of Tab.\ref{tab:lescas}, obtained in considering Mimas to be in hydrostatic equilibrium.

\subsection{Non-hydrostatic Mimas based on its shape}

\par As already mentioned, this case is unique, because changes in the size of the core do not affect the ratios of the moments of 
inertia $A/C$, and $B/C$, and the coefficients $\gamma_1$ and $\gamma_2$ (Eq.\ref{equ:gama1} and \ref{equ:gama2}). As a consequence, 
there is a unique rotational behavior of Mimas for any homogenous or 2-layer model using this specific model based on the observed 
shape.

\par The free librations around the equilibrium are assumed to be damped, it is anyway important to know their frequencies $\omega_u$, $\omega_v$ and $\omega_w$ (or periods $T_u$, $T_v$ and $T_w$) because they characterize the way the system reacts to external sinusoidal excitations, that are here due to the variations of the distance between the Sun and Mimas.

\begin{table}
\centering
\caption{Frequencies and periods of the free librations of Mimas, in the shape model. These values have been obtained numerically.\label{tab:freeshape}}
\begin{tabular}{l|rr}
\hline\hline
Proper & Frequency & Period $T$ \\
mode & (rad/d) & (d) \\
\hline
$u$ & $2.704622$ & $2.323129$ \\
$v$ & $0.778015$ & $8.075914$ \\
$w$ & $0.621287$ & $10.113182$ \\
\hline
\end{tabular}
\end{table}

\par The frequencies of the free librations are listed in Tab.\ref{tab:freeshape}. The proper mode $u$ roughly represents the free longitudinal librations, $v$ the free librations of the obliquity, and $w$ the wobble, i.e. the free polar motion of Mimas. These frequencies have been deduced from the frequency analysis of the modified Andoyer variables (cf.Eq.\ref{equ:modified}).

\par So, the proper modes involved in the Fourier representations of the librations of Mimas are the forced modes due to the orbital motion of Mimas around Saturn (cf.Tab.\ref{tab:propmod}) and the free ones (Tab.\ref{tab:freeshape}). The arguments of the sinusoidal components of the quasi-periodic decompositions of the variables of the rotation are integer combinations of these proper modes. If we consider that the free librations are damped, the solutions should be only composed of the forced modes.

\begin{table*}[ht]
\centering
\caption{Forced tidal longitudinal librations of Mimas, in the shape model. The series are in cosine.\label{tab:tidlongi}}
\begin{tabular}{rrrrrrrr}
\hline\hline
 &  & & & Frequency & Period & Amplitude & Phase \\
$\lambda$ & $\omega$ & $\phi$ & $\zeta$ & (rad/y) & (d) & (arcmin) & at J2000 \\
\hline
$1$ & -    & $-1$ & $1$ & $2428.763080$ & $0.944898$ & $157.73363$ & $-79.177^{\circ}$ \\
$1$ &  $1$ & $-1$ & $1$ & $2428.852395$ & $0.944863$ &   $5.72739$ & $-116.201^{\circ}$ \\
$1$ & $-1$ & $-1$ & $1$ & $2428.673643$ & $0.944933$ &   $4.05163$ & $137.568^{\circ}$ \\
$2$ & -    & $-2$ & $2$ & $4857.526150$ & $0.472449$ &   $1.83667$ & $-68.313^{\circ}$ \\
$1$ & -    & -    & $1$ & $2438.960801$ & $0.940947$ &   $1.32391$ & $-148.065^{\circ}$ \\
\hline
\end{tabular}
\end{table*}

\begin{table*}[ht]
\centering
\caption{Forced physical longitudinal librations of Mimas, in the shape model. The series are in cosine.\label{tab:physic}}
\begin{tabular}{rrrrrrrr}
\hline\hline
 &  & & & Frequency & Period & Amplitude & Phase \\
$\lambda$ & $\omega$ & $\phi$ & $\zeta$ & (rad/y) & (d) & (arcmin) & at J2000 \\
\hline
 -  & $1$ &  -   &  -  &    $0.08904538$ & $25772.62777$ & $43.61^{\circ}$ & $51.354^{\circ}$ \\
 -  & $3$ &  -   &  -  &    $0.26713614$ &  $8590.87592$ &  $43.261$ arcmin & $-25.913^{\circ}$ \\
$1$ &  -  & $-1$ & $1$ & $2428.763080$   &    $0.944898$ &  $26.075$ arcmin & $101.355^{\circ}$ \\
 -  &  -  &  $1$ &  -  &   $10.19765304$ &   $225.04526$ &   $7.828$ arcmin & $-157.744^{\circ}$ \\
 -  & $1$ & $-1$ &  -  &   $10.10860766$ &   $227.02728$ &   $3.657$ arcmin & $-119.032^{\circ}$ \\
 -  & $1$ &  $1$ &  -  &   $10.28669842$ &   $223.09718$ &   $3.532$ arcmin & $-16.309^{\circ}$ \\
\hline
\end{tabular}
\end{table*}

\par The forced librations of Mimas modeled from its observed shape are given in Tab.\ref{tab:tidlongi} to \ref{tab:obli}. These 
tables give the solutions under the form of periodic time series, in cosines. We can see that the main difference between the 
physical and the tidal librations is in the presence in the physical librations of a long-period term ($\approx70$ years) with a 
high amplitude ($\approx43^{\circ}$, i.e. $\approx86^{\circ}$ peak-to-peak) due to the librations of the argument of the orbital 
resonance between Mimas and Tethys. As explained in Rambaux \etal (2010) and (2011), the amplitude of the long period librations 
are equal to the magnitude of the orbital perturbations because at long period the body is oriented toward the central planet. 
As a consequence, by analysing the tidal librations, the long period librations vanish. There is also a large difference in the
amplitude given for the tidal and physical longitudinal librations. As explained above, this difference is due to optical librations,
with amplitude $2e\approx3.8\times10^{-2}$ rad $\approx2.2^{\circ}$.

\begin{table*}[ht]
\centering
\caption{Forced latitudinal librations of Mimas, in the shape model. The series are in cosine.\label{tab:lati}}
\begin{tabular}{rrrrrrrr}
\hline\hline
 & & & & Frequency & Period & Amplitude & Phase \\
$\lambda$ & $\omega$ & $\phi$ & $\zeta$ & (rad/y) & (d) & (arcmin) & at J2000 \\
\hline
$1$ & -    & $1$ & $-1$     & $2441.516177$ &   $0.939962$ & $2.07096$ & $77.130^{\circ}$ \\
$1$ & $1$  & $1$ & $-1$     & $2441.605507$ &   $0.939928$ & $0.06829$ & $39.984^{\circ}$ \\
$1$ & $-1$ & $1$ & $-1$     & $2441.426665$ &   $0.939997$ & $0.0414$ & $-66.603^{\circ}$ \\
\hline
\end{tabular}
\end{table*}

\begin{table*}[ht]
\centering
\caption{Forced obliquity of Mimas, in the shape model. The series are in cosine.\label{tab:obli}}
\begin{tabular}{rrrrrrrr}
\hline\hline
 &  & & & Frequency & Period & Amplitude & Phase \\
$\lambda$ & $\omega$ & $\phi$ & $\Phi$ & (rad/y) & (d) & (arcmin) & at J2000 \\
\hline
 -  & -    &  -      & -    & $0$           &   $\infty$ & $2.13468$ & - \\
$2$ & -    & $1$     & $-1$ & $4883.032354$ & $0.469981$ & $0.07372$ & $-27.144^{\circ}$ \\
$2$ & $-1$ & $1$     & $-1$ & $4882.943311$ & $0.469990$ & $0.07179$ & $14.019^{\circ}$ \\
$2$ & $1$  & $1$     & $-1$ & $4883.121227$ & $0.469973$ & $0.06866$ & $119.800^{\circ}$ \\
-   & $2$  & -       & -    &    $0.177998$ & $12893.06$ & $0.04817$ & $103.293^{\circ}$ \\
\hline
\end{tabular}
\end{table*}

\par The latitudinal librations of Mimas (Tab.\ref{tab:lati}) are significantly smaller ($\approx2$ arcmin vs. $2.5^{\circ}$ for the tidal longitudinal librations), and so could hardly be used in the framework of observations of the rotation of Mimas (except if there are free oscillations due to a recent unexpected excitation). The mean obliquity of Mimas (Tab.\ref{tab:obli}) is of the same order of magnitude.

\subsection{For a hydrostatic Mimas}

\par We performed the same numerical study of the 22 hydrostatic configurations of Mimas given in Tab.\ref{tab:lescas}. The results are gathered in Tab.\ref{tab:reshydro}. In this table, the amplitudes of the tidal and physical longitudinal librations indicated are related to the mode $\lambda-\phi+\zeta$ (period: 0.944898 d), while the latitudinal ones are related to the mode $\lambda+\phi-\zeta$ (period: 0.939962 d). The main physical reason of these librations is the variations of the distance Mimas-Saturn during an orbital period.

\begin{table*}[ht]
\centering
\caption{Periods of the free librations and amplitudes (in arcmin) of the forced librations for the different models assuming that Mimas is at the hydrostatic equilibrium.\label{tab:reshydro}}
\begin{tabular}{l|rrrrrrr}
\hline\hline
 & $T_u$ & $T_v$ & $T_w$ & Tidal & Latitudinal & Mean & Physical \\
N & (d) & (d) & (d) & librations & librations & obliquity & librations \\
\hline
 1 & $2.143878$ &  $7.885550$ & $11.621674$ & $163.398$ & $2.016$ & $2.086$ & $31.744$ \\
 2 & $2.294081$ &  $8.997072$ & $13.222674$ & $158.577$ & $2.314$ & $2.384$ & $26.914$ \\
 3 & $2.407777$ &  $9.908107$ & $12.763086$ & $155.609$ & $2.559$ & $2.631$ & $23.944$ \\
 4 & $2.468518$ & $10.416095$ & $13.618627$ & $154.248$ & $2.693$ & $2.765$ & $22.582$ \\
 5 & $2.507237$ & $10.742236$ & $14.181571$ & $153.442$ & $2.780$ & $2.853$ & $21.776$ \\
 6 & $2.534519$ & $10.975456$ & $14.591169$ & $152.900$ & $2.843$ & $2.917$ & $21.234$ \\
 7 & $2.555063$ & $11.152890$ & $14.906138$ & $152.508$ & $2.891$ & $2.966$ & $20.841$ \\
\hline 
 8 & $2.138844$ &  $7.849766$ & $11.569966$ & $163.583$ & $2.006$ & $2.076$ & $31.922$ \\
 9 & $2.220696$ &  $8.443608$ & $12.426948$ & $160.777$ & $2.166$ & $2.234$ & $29.115$ \\
10 & $2.257477$ &  $8.730512$ & $12.839826$ & $159.593$ & $2.243$ & $2.313$ & $27.930$ \\
11 & $2.274983$ &  $8.851150$ & $13.013162$ & $159.124$ & $2.275$ & $2.346$ & $27.461$ \\
12 & $2.284054$ &  $8.920304$ & $13.112486$ & $158.862$ & $2.293$ & $2.363$ & $27.199$ \\
13 & $2.290031$ &  $8.966015$ & $13.178106$ & $158.692$ & $2.306$ & $2.378$ & $27.028$ \\
14 & $2.294326$ &  $8.998974$ & $13.225425$ & $158.571$ & $2.315$ & $2.384$ & $26.907$ \\
\hline
15 & $2.128297$ &  $7.775062$ & $11.461869$ & $163.975$ & $1.986$ & $2.054$ & $32.314$ \\
16 & $2.154684$ &  $7.962727$ & $11.733240$ & $163.008$ & $2.037$ & $2.104$ & $31.347$ \\
17 & $2.162067$ &  $8.015705$ & $11.809777$ & $162.745$ & $2.051$ & $2.119$ & $31.084$ \\
18 & $2.164709$ &  $8.034707$ & $11.837202$ & $162.652$ & $2.056$ & $2.124$ & $30.991$ \\
19 & $2.166135$ &  $8.044989$ & $11.852091$ & $162.602$ & $2.059$ & $2.127$ & $30.941$ \\
20 & $2.167044$ &  $8.051527$ & $11.861492$ & $162.570$ & $2.060$ & $2.132$ & $30.909$ \\
21 & $2.167685$ &  $8.056158$ & $11.868221$ & $162.548$ & $2.062$ & $2.130$ & $30.886$ \\
\hline
22 & $2.106951$ &  $7.625231$ & $11.244996$ & $164.792$ & $1.946$ & $2.014$ & $33.132$ \\
\hline
 \end{tabular}
\end{table*}

\par
\begin{figure*}
\centering
\begin{tabular}{cc}
\includegraphics[width=6.3cm,height=4.5cm]{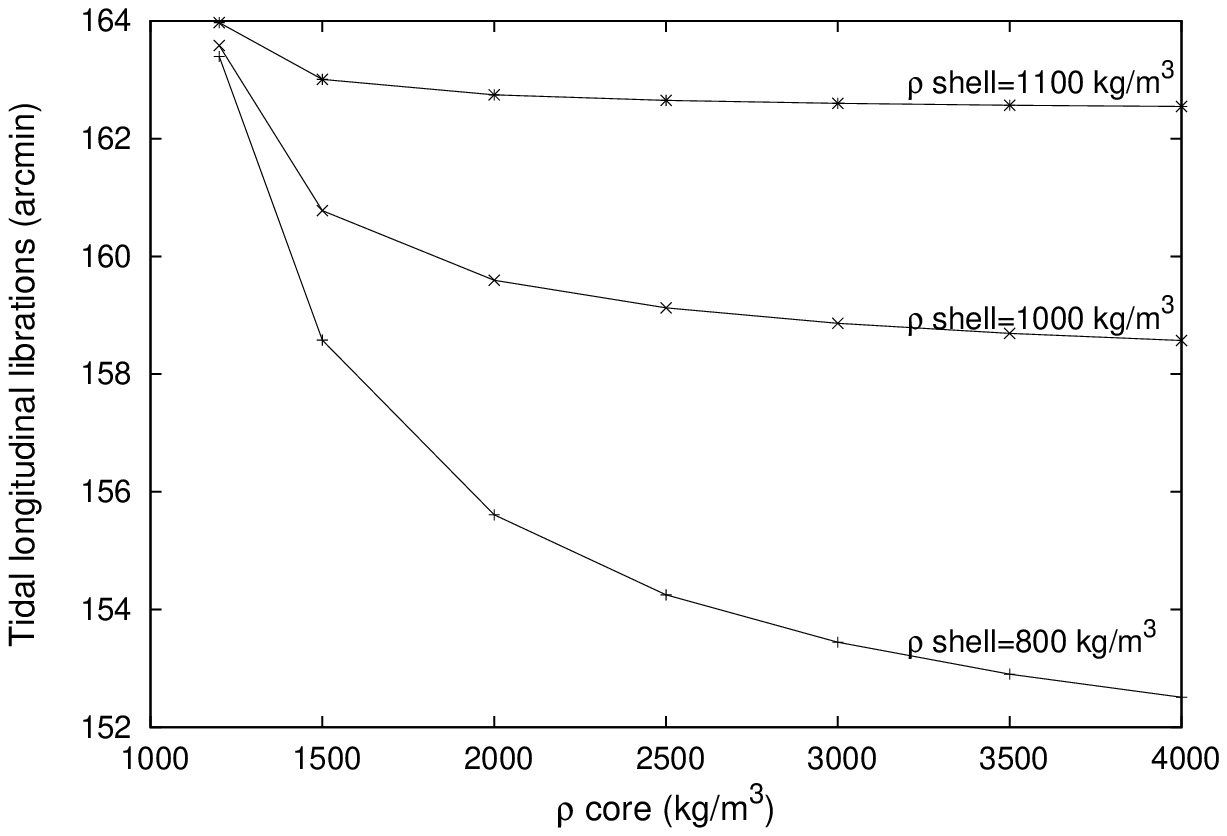} & \includegraphics[width=6.3cm,height=4.5cm]{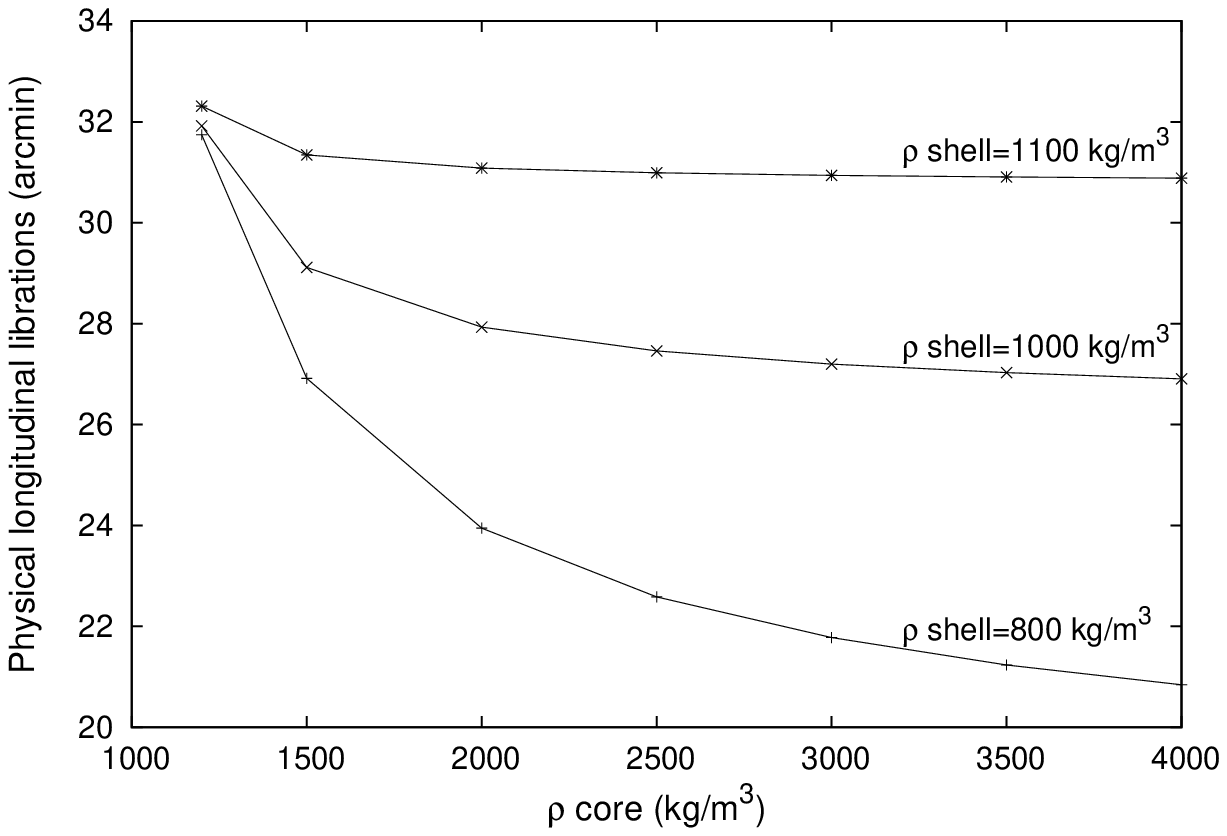} \\
\includegraphics[width=6.3cm,height=4.5cm]{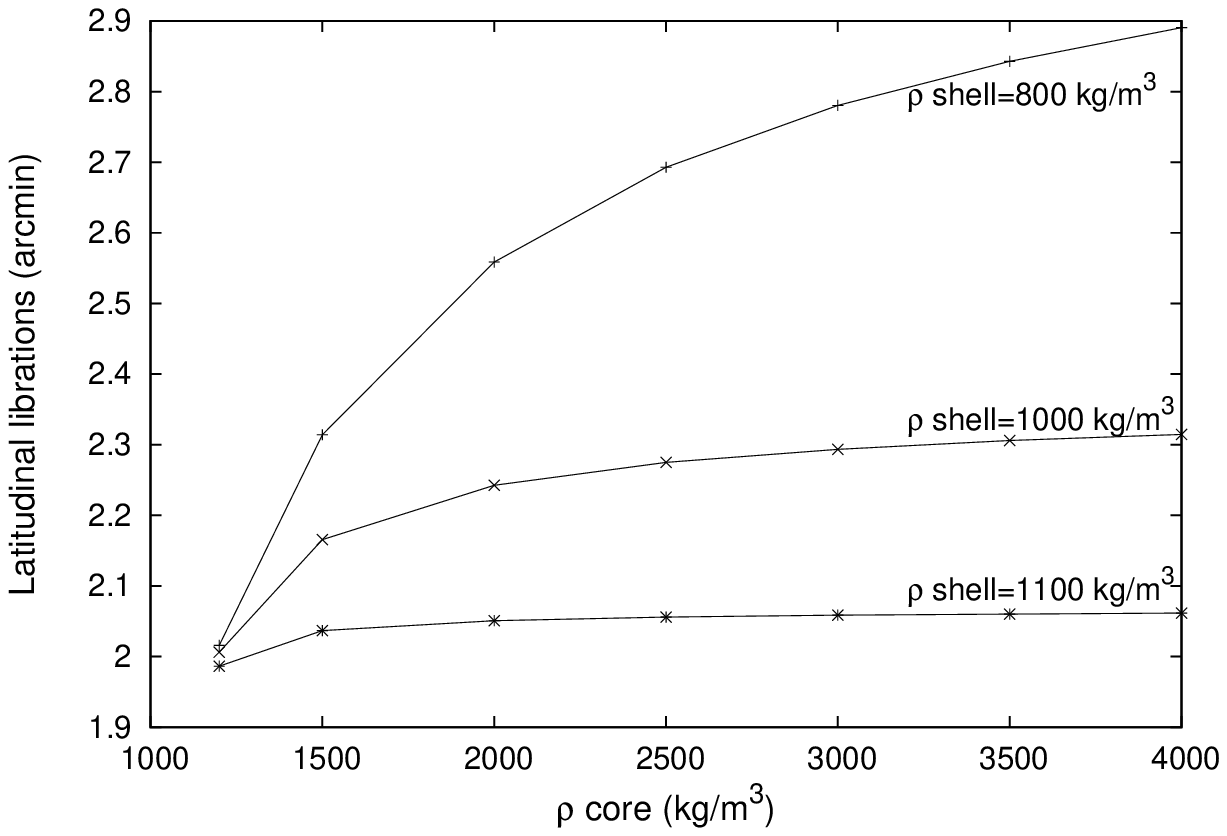} & \includegraphics[width=6.3cm,height=4.5cm]{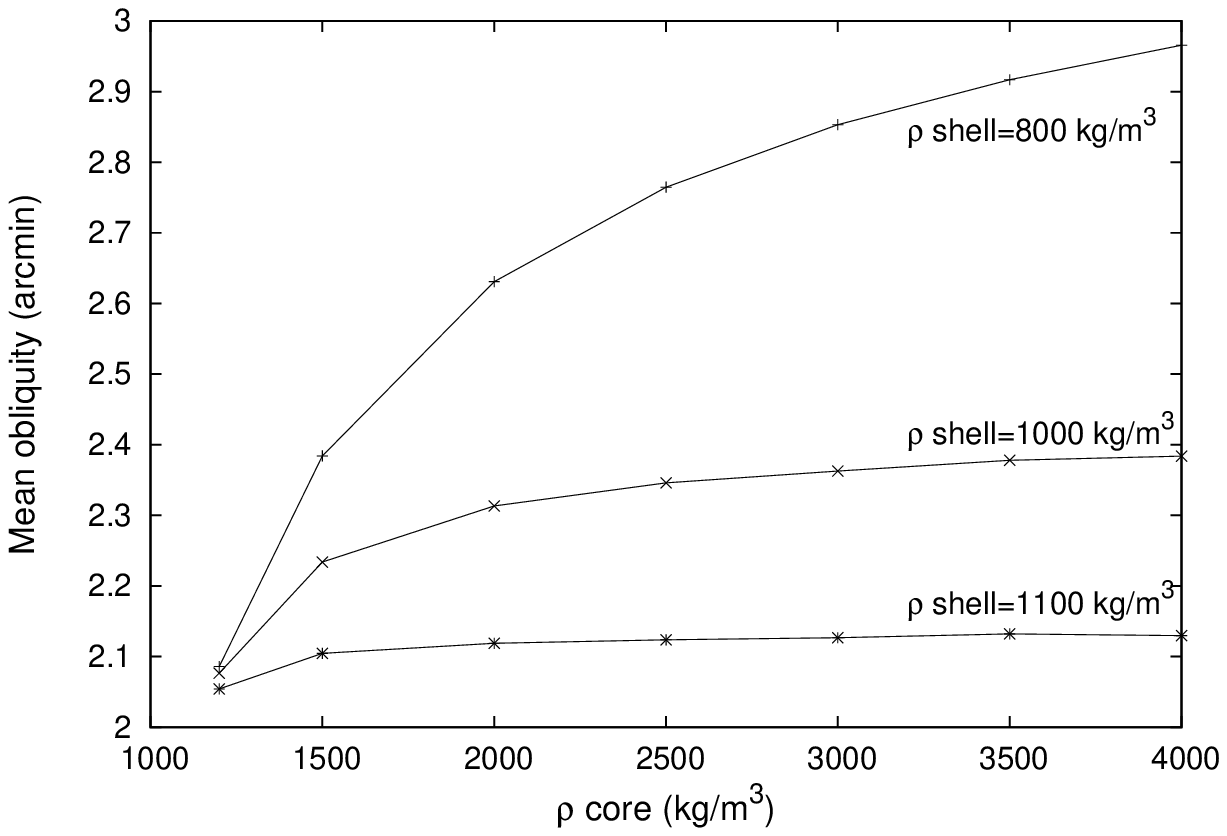}
\end{tabular}
\caption{Librations of hydrostatic Mimas, depending on the densities of the core and the shell.\label{fig:librations}}
\end{figure*}

 In order to make the results more readable, we present them graphically in Fig.\ref{fig:librations}. The plots present a clear dependency of the amplitudes of librations on the densities of the core and the shell. We can in particular notice that the longitudinal (i.e. tidal and physical) librations have a larger amplitude when the density of the core is lower, it is due to the fact that a concentration of the mass in the core lowers the moments of inertia of the body, and so tends to limit its amplitude of response to sollicitations. Finally we can see that the dependency on $\rho_c$ is small for $\rho_s=1100kg/m^3$, it is because in this case, $\rho_s$ is close to the mean density of Mimas (i.e. $1150.03kg/m^3$), as a consequence the core is small and Mimas is close to be homogeneous.

\subsection{A small polar motion}

\par As for the other outputs, we present the forced polar motion of Mimas (i.e. after removal of the free wobble) as a sum of a 
trigonometric series (Tab.\ref{tab:polar}). We can see that this motion is expected to be small, the highest amplitude being 
$\approx15$ meters. The sum of all these amplitudes can reach 40 meters, so we can consider these 40 meters as the upper bound of 
the polar motion. An analysis of the polar motions for the different hydrostatic Mimas do not exhibit significant differences.

\begin{table*}[ht]
\centering
\caption{Polar motion of Mimas $q_1+\imath q_2$, in the shape model. The series are in complex exponentials.\label{tab:polar}}
\begin{tabular}{rrrrrrrrr}
\hline\hline
 &  & & & & Frequency & Period & Amplitude & Phase \\
$\lambda$ & $\omega$ & $\phi$ & $\zeta$ & $\Phi$ & (rad/y) & (d) & (m) & at J2000 \\
\hline
$1$  & -    & $1$  & $-1$ & -    & $2441.516177$  &   $0.939962$ & $15.277$ & $77.130^{\circ}$ \\
$-1$ & -    & $-1$ & $1$  & -    & $-2441.516177$ &   $0.939962$ & $14.368$ & $102.870^{\circ}$ \\
 -   & -    & $-1$ & -    &  $1$ & $-12.7530964$  & $179.951077$ &  $4.498$ & $113.654^{\circ}$ \\
 -   & -    & $1$  & -    & $-1$ & $12.7530964$   & $179.951077$ &  $3.441$ & $66.208^{\circ}$ \\
$1$  & $1$  & $1$  & $-1$ & -    & $2441.605507$  &   $0.940313$ &  $0.503$ & $40.052^{\circ}$ \\
$2$  & -    & $-1$ &  $2$ & $-1$ & $4870.2792464$ &   $0.471212$ &  $0.485$ & $87.878^{\circ}$ \\
$-1$ & $-1$ & $-1$ & $1$  & -    & $-2441.605507$ &   $0.940313$ &  $0.473$ & $139.948^{\circ}$ \\
\hline
\end{tabular}
\end{table*}

\section{Tidal Dissipation}\label{sec:tidal}

This section is dedicated to study the influence of the tidal torque on the rotational motion of Mimas. We introduce the tidal torque in a Lagrangian formalism and follow the approach of Williams \etal (2001) and used recently in Rambaux \etal (2010) and Robutel \etal (2011). The starting equation is the angular momentum equation
\begin{equation}
\frac{d \vec{G}}{dt} + \vec{\omega} \wedge \vec{G}= \vec{T}
\label{eq:lag}
\end{equation}
where $\vec{\omega}$ is the angular velocity vector, the angular momentum $\vec G=I \vec \omega$ with $I$ the tensor of inertia, and $\vec{T}$ is the external gravitational torque expressed as 
\begin{equation}
\vec T = \frac{3GM_{\saturn}}{r^3} \vec u \wedge I \vec u  
\end{equation}
where $\vec u$ is the cosine director of Saturn in the reference frame tied to Mimas, and $M_{\saturn}$ its mass.

The dissipation is due to the tidal and centrifugal potentials that deform the satellite. In this case, the tensor of inertia $I$ 
becomes a constant plus a time-variable part resulting from the deformation. The time-variable part does not react instantaneously 
and therefore presents a time delay $\delta t$ characteristics of the rheological properties of the body (see section~\ref{sec:is}).    

In addition, the dynamical equation of the rotational motion Eq.~\ref{eq:lag} may be linearized by using the synchronous spin-orbit 
resonance of the body implying that $ \omega_1, \omega_2 << \omega_3 \sim n $ and $u_2, u_3 << u_1 \sim 1 $ where $u_1, u_2, u_3$ are the coordinates of the cosine director 
along the principal axis of inertia of Mimas.

By using these approximations and focusing on the libration in longitude, the main tidal torque is expressed as (Williams \etal 2001)
\begin{equation}
T_t = -k_2 R^5 \frac{3 GM_{\saturn}^2}{a^6}( U_{11}U^*_{12} - U_{12}U^*_{11}), 
\end{equation}
where $U_{ij}=\left(\frac{a}{r}\right)^3 u_i u_j$ and the star indicates the time delay part.

Then, we used the same approach than in Rambaux \etal (2010). We introduce the rotation angle $\varphi$ similar to the sum of the 
Andoyer angles $l+g$ because the polar motion is small as shown in Figure~\ref{fig01} where $J$ and $\theta$ (the nutation angle) 
are small. The libration angle $\gamma$ is defined as $\varphi = M+\gamma $ representing the physical libration in longitude of the 
body. The cosine director $u_2 \sim s-\gamma $ is of the order of the difference between the orbital variation $s$ and the physical 
libration $\gamma$. We note that $u_2$ corresponds to the tidal libration $\psi$ defined in section~\ref{sec:output} and their 
amplitude is small as shown in Table 5. The quantity $s$, the orbital variation, is defined as the difference between the true and 
the mean longitude of Mimas and represents the oscillation of the orbital longitude of Mimas that may be expressed in Fourier series 
as 
\begin{equation}
s = \sum_i H_i \sin{(\omega_i t + \alpha_i)}.
\end{equation}
Then, by developing $u_1^*$ and $u_2^*$ in Taylor series for small $\delta t$, the dynamical equation becomes:
\begin{equation}
C \ddot \gamma + \frac{3}{2} (B-A) \frac{Gm}{r^3} \sin{2(\gamma - s)} =  -k_2 R^5 \frac{3 GM_{\saturn}^2}{a^6} \delta t ( \dot{\gamma} - \dot{s} ).
\label{eq:gammalindissip}
\end{equation}
As shown in the previous section, the quantity $\gamma -s$ is always small (see Table 5) allowing to simplify the sine function by its angle. In addition, the eccentricity of Mimas is small and so $a/r$ is equal to 1 at first order in eccentricity. Finally, we obtain a forced dissipative harmonic oscillator with the frequency $\omega_0 = n \sqrt{3 (B-A)/C}$ and the dissipative rate $\lambda$ expressed as 
\begin{equation}
2 \lambda = \frac{3 k_2 R^3}{C} \frac{n^4}{Gm} \delta t,
\end{equation}
$m$ being the mass of the satellite and $k_2$ is the Love number of Mimas.

As for the conservative case, the amplitude of terms associated with the long period is almost equal to the magnitude of the oscillation $s$. The solution may be expressed as 
\begin{equation}
\gamma = A_d\sin{(\omega_d t + \phi_d)}e^{-\lambda t} + \sum_i  x_i \cos{(\omega_i t+ \alpha_i )} + y_i \sin{(\omega_i t+ \alpha_i )},
\end{equation}
where $A_d$ and $\phi_d$ are constants of integration. The first term decays with time scale $1/\lambda$ and its resonant frequency is $\omega_d = \sqrt{\omega_0^2 - \lambda^2}$. The periodic term of the particular solution $\gamma$ is composed of the in-phase $y_i$ and out-of-phase $x_i$ terms  
\begin{equation}
y_i  = H_i  \frac{ (\omega_0^2-\omega_i^2)\omega_0^2+4\lambda^2 \omega_i^2 }{(\omega_0^2-\omega_i^2)^2+4\lambda^2 \omega_i^2} \;\;\;  , \;\;\;
x_i  =  H_i \frac{-2\lambda \omega_i^3}{(\omega_0^2-\omega_i^2)^2+4\lambda^2 \omega_i^2}.
\label{phaseshift}
\end{equation}

At first order, the expression of $x_i$ may be simplified as 
\begin{equation}
x_i  = -0.9054 \frac{k_2}{Q} H_i \frac{\omega_i^3}{(\omega_0^2-\omega_i^2)^2} 
\end{equation}
expressed in radians and by introducing the dissipation factor as $\delta t = (n Q)^{-1}$.
For short period librations at 0.944898 days the $x_i$ is 1.32 mas with $\frac{k_2}{Q} = 10^{-6}$ (this is the value used by 
\cite{mw08}) and the resulting displacement at the surface of the satellite at the periaster passage is also negligible 0.0013 m. 
The damping time $1/\lambda$ is about 6,000 years. If we consider $\frac{k_2}{Q}$ to be 100 times bigger, i.e. closer to the 
expected value of Enceladus, we have a displacement at the surface of $\approx0.13$ m and a damping time of $\approx60$ years. 
For librations at long period $x_i$ is definitely negligible because $\omega_i$ is very small.

\section{Discussion}

\par One of the aims of this theoretical study is to prepare the interpretation of potential observations of the rotation of Mimas. After a restricted analytical approach to validate the numerical results, we discuss the possibility to observe the rotation of Mimas and in particular to discriminate the different interior models. Then we focus on the non-hydrostatic contributions.

\subsection{Analytical approach}

\par We here compare with classical analytical formulae for the main term of the physical and tidal longitudinal librations and the mean obliquity, for which deriving accurately these amplitudes is quite straightforward.

\subsubsection{Longitudinal librations}

\par An analytical study of the longitudinal librations of a satellite in 1:1 spin-orbit resonance and on a keplerian orbit can be 
found for instance in \cite{md99}. Let us call $\psi$ the amplitude of the main term (i.e. associated with the 
mode $\lambda+\phi-\zeta$) of the tidal librations, and $\gamma$ for the physical ones. We have from \cite{md99}:

\begin{eqnarray}
\psi & = & \frac{-2e}{1-\left(\frac{\omega_u}{n}\right)^2}, \label{eq:psi} \\
\gamma & = & \frac{2e}{1-\left(\frac{n}{\omega_u}\right)^2}, \label{eq:gamma} \\
\end{eqnarray}
and 

\begin{equation}
\label{eq:omegu}
\left(\frac{\omega_u}{n}\right)^2=3\frac{B-A}{C}\left(1-5e^2+\frac{13}{16}e^4\right)=12\frac{C_{22}}{C/(mR^2)}\left(1-5e^2+\frac{13}{16}e^4\right),
\end{equation}
$e$ being the eccentricity of Mimas. We can see that this amplitude is bigger when the ratio is closer to unity, or when $12C_{22}$ is closer to $C/(mR^2)$. We can see from the Tab.\ref{tab:lescas} that $C_{22}$ is of the order $5\times10^{-3}$ while $C\approx0.4mR^2$, i.e. $C_{22}/C/(mR^2)\approx1/80$. Thus, the ratio $12\frac{C_{22}}{C/(mR^2)}$ is closer to unity when $C_{22}$ is bigger, what is the case for the smallest values of $\rho_c$. The Fig.\ref{fig:librations} confirms this trend, while the Tab.\ref{tab:confirms} settles the validity of the analytical formulae (\ref{eq:psi}) and (\ref{eq:gamma}).

\subsubsection{Mean obliquity}

\par We here use the analytical study of \cite{wh04} (see \cite{n10} for an application to natural satellites in spin-orbit resonances) for the location of the Cassini States. Mimas is expected to be locked at the Cassini State 1, i.e. the most stable one, characterized by:

\begin{equation}
\label{eq:ward}
\epsilon=-\frac{\sin I}{\frac{3n}{2\dot{\ascnode}}\frac{J_2+2C_{22}}{C/(mR^2)}+\cos I},
\end{equation}
$\epsilon$ being the mean obliquity of Mimas, $\dot{\ascnode}$ the precessional rate of its orbital ascending node, and $I$ its inclination on the Laplace Plane, here assumed to be the equator of Saturn at J2000.

\par From the definition the orbital proper modes of Mimas, we can approximate $\dot{\ascnode}$ by $\dot{\Phi}-\dot{\zeta}=-6.37188169$ 
rad/y, this yields a regressional period of $360.1657$ days. In assuming $J_2\approx2\times10^{-2}$, $C_{22}\approx6\times10^{-3}$ 
and $C\approx0.4mR^2$ from Tab.\ref{tab:lescas}, we have $\frac{3n}{2\dot{\ascnode}}\frac{J_2+2C_{22}}{C/(mR^2)}\approx-37.26$ 
while $\sin I$ is very small and $\cos I$ close to unity (the mean inclination of Mimas $I$ being of the order of 
$1.5^{\circ}=2.6\times10^{-2}$ rad). So, bigger values of the quantity $J_2+2C_{22}$ will yield a smaller obliquity. Once more, these values are reached for the lowest estimations of $\rho_c$, the Fig.\ref{fig:librations} confirming this tendency. The validity of the analytical formula (\ref{eq:ward}) is checked in Tab.\ref{tab:confirms}.

\begin{table*}[ht]
\centering
\caption{Analytical confirmation of the numerical results given in Tab.\ref{tab:lescas}. The analytical formulae used are Eq.\ref{eq:omegu}, \ref{eq:gamma}, \ref{eq:psi} \& \ref{eq:ward}, the obtained values being compared with the ones given in Tab.\ref{tab:reshydro}.\label{tab:confirms}}
\begin{tabular}{l|rrrrrrrr}
\hline\hline
  & $T_u$ & $\Delta T_u$ & $\gamma$   & $\Delta\gamma$ & $\psi$      & $\Delta\psi$ &  $\epsilon$     & $\Delta\epsilon$ \\
N & (d)  &               & (arcmin) &                & (arcmin) &              & (arcmin) & \\
\hline
 1 & $2.145171$ & $0.060\%$ & $31.572$ & $0.542\%$ & $163.581$ & $0.112\%$ & $2.088$ & $0.104\%$ \\
 2 & $2.294910$ & $0.036\%$ & $26.778$ & $0.506\%$ & $158.787$ & $0.133\%$ & $2.397$ & $0.547\%$ \\
 3 & $2.410238$ & $0.102\%$ & $23.825$ & $0.497\%$ & $155.835$ & $0.145\%$ & $2.651$ & $0.741\%$ \\
 4 & $2.471008$ & $0.101\%$ & $22.471$ & $0.493\%$ & $154.480$ & $0.150\%$ & $2.790$ & $0.890\%$ \\
 5 & $2.509751$ & $0.100\%$ & $21.669$ & $0.490\%$ & $153.679$ & $0.154\%$ & $2.880$ & $0.956\%$ \\
 6 & $2.537080$ & $0.101\%$ & $21.131$ & $0.487\%$ & $153.140$ & $0.157\%$ & $2.945$ & $0.966\%$ \\
 7 & $2.557614$ & $0.100\%$ & $20.740$ & $0.487\%$ & $152.749$ & $0.158\%$ & $2.994$ & $0.960\%$ \\
\hline 
 8 & $2.140163$ & $0.062\%$ & $31.755$ & $0.522\%$ & $163.765$ & $0.111\%$ & $2.078$ & $0.108\%$ \\
 9 & $2.221696$ & $0.045\%$ & $28.965$ & $0.514\%$ & $160.975$ & $0.123\%$ & $2.243$ & $0.411\%$ \\
10 & $2.259963$ & $0.110\%$ & $27.788$ & $0.508\%$ & $159.798$ & $0.128\%$ & $2.323$ & $0.429\%$ \\
11 & $2.275851$ & $0.038\%$ & $27.321$ & $0.508\%$ & $159.331$ & $0.130\%$ & $2.356$ & $0.446\%$ \\
12 & $2.284904$ & $0.037\%$ & $27.061$ & $0.507\%$ & $159.071$ & $0.131\%$ & $2.376$ & $0.537\%$ \\
13 & $2.290872$ & $0.037\%$ & $26.892$ & $0.505\%$ & $158.901$ & $0.132\%$ & $2.388$ & $0.438\%$ \\
14 & $2.295156$ & $0.036\%$ & $26.771$ & $0.506\%$ & $158.780$ & $0.132\%$ & $2.398$ & $0.569\%$ \\
\hline
15 & $2.129674$ & $0.065\%$ & $32.145$ & $0.522\%$ & $164.155$ & $0.110\%$ & $2.057$ & $0.170\%$ \\
16 & $2.155925$ & $0.058\%$ & $31.184$ & $0.522\%$ & $163.193$ & $0.113\%$ & $2.110$ & $0.266\%$ \\
17 & $2.163275$ & $0.056\%$ & $30.922$ & $0.520\%$ & $162.932$ & $0.115\%$ & $2.124$ & $0.251\%$ \\
18 & $2.165905$ & $0.055\%$ & $30.830$ & $0.520\%$ & $162.839$ & $0.115\%$ & $2.130$ & $0.263\%$ \\
19 & $2.167325$ & $0.055\%$ & $30.780$ & $0.520\%$ & $162.789$ & $0.115\%$ & $2.132$ & $0.256\%$ \\
20 & $2.168230$ & $0.055\%$ & $30.748$ & $0.520\%$ & $162.758$ & $0.116\%$ & $2.134$ & $0.106\%$ \\
21 & $2.168868$ & $0.055\%$ & $30.726$ & $0.518\%$ & $162.736$ & $0.115\%$ & $2.136$ & $0.261\%$ \\
\hline
22 & $2.108464$ & $0.074\%$ & $32.958$ & $0.526\%$ & $164.967$ & $0.106\%$ & $2.016$ & $0.095\%$ \\
\hline
 \end{tabular}
\end{table*}

\par The analytical validation meets the following trouble: how to evaluate the mean eccentricity and inclination required in the analytical formulae, i.e. how to average them? These formulae have been derived in assuming a Keplerian orbit, while the orbit of Mimas is perturbed by the oblateness of Saturn and the mutual interactions with the other satellites, inducing an orbital resonance with Tethys. As a consequence, its eccentricity and inclination are far from constant.

\par We have, from \cite{vd95}:

\begin{eqnarray}
z(t) & = & e(t)\exp\left(\imath\varpi(t)\right) \nonumber \\
& = & 1.59817\times10^{-2}\exp\left(\imath\left(6.38121472t+356.521^{\circ}\right)\right) \nonumber \\
& + & 7.2147\times10^{-3}\exp\left(\imath\left(6.29216934t+137.197^{\circ}\right)\right)  \label{eq:ztass}  \\
& + & 7.1114\times10^{-3}\exp\left(\imath\left(6.47026010t+35.846^{\circ}\right)\right)+\ldots, \nonumber \\
\zeta(t) & = & \sin\left(\frac{I(t)}{2}\right)\left(\imath\ascnode(t)\right) \nonumber \\
& = & 1.18896\times10^{-2}\exp\left(\imath\left(-6.37188169t+234.213^{\circ}\right)\right) \nonumber \\
& + & 5.3177\times10^{-3}\exp\left(\imath\left(-6.46092707t+14.888^{\circ}\right)\right) \label{eq:zetatass} \\
& + & 5.3017\times10^{-3}\exp\left(\imath\left(-6.28283631t+273.538^{\circ}\right)\right)+\ldots, \nonumber
\end{eqnarray}
the frequencies being in rad/year, and the time origin J1980. As we can see, the mean eccentricity should be at least $\approx1.6\times10^{-2}$, probably higher (same for the mean inclination, that should be at least $\approx1.4^{\circ}$). In the Tab.\ref{tab:confirms}, we use $e=1.92\times10^{-2}$ and $I=1.68^{\circ}$, this arbitrary choice minimizes the relative errors and is consistent with the TASS1.6 theory.

\subsection{Observational possibilities}

\par It would be challenging to constrain the orientation and interior structure of Mimas using its rotation. The first expected 
result is the confirmation that Mimas is in the Cassini State 1 with the 1:1 spin-orbit resonance. Another challenge would be to 
detect the longitudinal librations, that have been actually observed for the Moon (\cite{k67}), the Martian satellite 
Phobos (\cite{b72}), and the Saturnian satellite Epimetheus (\cite{ttb09}). To estimate the required accuracy of the observations, we convert the rotation outputs into kilometres (Tab.\ref{tab:km}).

\begin{table}[ht]
\centering
\caption{Expected librations and mean obliquity of Mimas, in km. The mean obliquity $\epsilon$ has been multiplied by the polar radius $c=190.6$ km, while the librations have been multiplied by the Saturn-facing radius $a=207.8$ km. The case 23 is derived from the shape model. \label{tab:km}}
\begin{tabular}{l|rrrrr}
\hline\hline
  & Physical   & Tidal      & Latitudinal & Mean \\
N & librations & librations & librations  & Obliquity \\
\hline
 1 & $1.919$ & $9.877$ & $0.122$ & $0.116$ \\
 2 & $1.627$ & $9.585$ & $0.140$ & $0.132$ \\
 3 & $1.447$ & $9.406$ & $0.155$ & $0.146$ \\
 4 & $1.365$ & $9.324$ & $0.163$ & $0.153$ \\
 5 & $1.316$ & $9.275$ & $0.168$ & $0.158$ \\
 6 & $1.284$ & $9.242$ & $0.172$ & $0.162$ \\
 7 & $1.260$ & $9.219$ & $0.175$ & $0.164$ \\
\hline
 8 & $1.930$ & $9.888$ & $0.121$ & $0.115$ \\
 9 & $1.760$ & $9.718$ & $0.131$ & $0.124$ \\
10 & $1.688$ & $9.647$ & $0.136$ & $0.128$ \\
11 & $1.660$ & $9.618$ & $0.138$ & $0.130$ \\
12 & $1.644$ & $9.603$ & $0.139$ & $0.131$ \\
13 & $1.634$ & $9.592$ & $0.139$ & $0.132$ \\
14 & $1.626$ & $9.585$ & $0.140$ & $0.132$ \\
\hline
15 & $1.953$ & $9.912$ & $0.120$ & $0.114$ \\
16 & $1.895$ & $9.853$ & $0.123$ & $0.117$ \\
17 & $1.879$ & $9.837$ & $0.124$ & $0.117$ \\
18 & $1.873$ & $9.832$ & $0.124$ & $0.118$ \\
19 & $1.870$ & $9.829$ & $0.124$ & $0.118$ \\
20 & $1.868$ & $9.827$ & $0.125$ & $0.118$ \\
21 & $1.867$ & $9.825$ & $0.125$ & $0.118$ \\
\hline
22 & $2.003$ & $9.961$ & $0.118$ & $0.112$ \\
\hline
23 & $1.576$ & $9.534$ & $0.125$ & $0.118$ \\
\hline
\end{tabular}
\end{table}

\par As expected, the longitudinal librations are significantly bigger (a few kilometres) than the mean obliquity and the latitudinal librations (with an amplitude smaller than 200 meters). The amplitude of the librations given are related to the quasi-periodic decompositions, so the peak-to-peak amplitudes are twice bigger. The reader should keep in mind that the physical and tidal librations are two expressions of the same quantity, so are not independent. We can consider that the detection of the longitudinal librations would require an accuracy of about 1 km, while using them to invert the internal structure of Mimas would require an accuracy at least ten times better.

\subsection{Non-hydrostatic contributions}

\par The study of the non-hydrostatic Mimas, based on the shape model, does not exhibit a significant possibility to discriminate a 
non-hydrostatic Mimas from a hydrostatic one from observations. This is not surprising considering Mimas' nearly hydrostatic global 
shape. But a non-hydrostatic Mimas could result in an offset between the 
ellipsoid of shape and the ellipsoid of inertia, as investigated for Janus by \cite{rrc11}, for which an offset in longitude and 
in latitude has actually been detected (\cite{ttb09}). So, detection of non-hydrostatic contributions from observation of 
Mimas' orientation should not a priori be excluded.

\section{Conclusion}

\par We have presented a theoretical study of the rotation of Mimas, in considering the 3 degrees of freedom of the rigid rotation, and different possible interior models, in assuming Mimas to be in hydrostatic equilibrium, or not. Moreover, we have considered a complete orbital motion, and also investigated the influence of tides on the rotation of Mimas.

\par We estimate the physical longitudinal librations to have an amplitude of about $0.5^{\circ}$, i.e. nearly 2 km, the exact 
value depending on the internal structure of Mimas. For a hydrostatic Mimas, a dense core lowers this amplitude. Non-hydrostatic 
contributions are shown to be small as expected from Mimas shape in near hydrostatic equilibrium. Moreover, we 
expect an obliquity between 2 and 3 arcmin, while the polar motion can be neglected. The tidal deviation of Mimas' long axis 
should be negligible as well, while this is the most inner main Saturnian satellite.

\par The Cassini spacecraft has already completed its initial four-year mission and the first extended mission, with a limited 
number of Mimas flybys. Its orbit close to Saturn makes Mimas a difficult target for Cassini observations. Since September 2010 
Cassini is in a second extended mission called the Cassini Solstice Mission during (and especially at the end) of which Cassini 
will likely have additional Mimas observations.  We hope that future observations of Mimas will allow us to constrain its rotation 
and to get clues on its internal structure and orientation.

\appendix

\section{Notations used in the paper}

\begin{table*}[ht]
\centering
\caption{Notations used in the paper.\label{tab:nota}}
\begin{tabular}{ll}
\hline\hline
\multicolumn{2}{c}{Physical parameters} \\
\hline
$R$, $R_c$ & Mean radius of Mimas and of its core \\
$m$ & Mass of Mimas \\
$a>b>c$ & Radii of Mimas \\
$A<B<C$ & Moments of inertia of Mimas \\
$A_c<B_c<C_c$ & Moments of inertia of the core of Mimas \\
$\rho_c>\rho_s$ & Densities of the core and the shell of Mimas \\
$k_f$ & Fluid Love number of Mimas \\
$k_2$ & Love number of Mimas \\
$J_2=-C_{20}$, $C_{22}$ & Gravity coefficients of Mimas \\
\hline
\multicolumn{2}{c}{Proper modes} \\
\hline
$\lambda$, $\omega$, $\phi$, $\zeta$, $\Phi$ & Orbital modes \\
$n$ & Orbital and spin frequency of Mimas \\
$u$, $v$, $w$ & Rotational modes, respectively in longitude, latitude and wobble\\
$\omega_u$, $\omega_v$, $\omega_w$ & Frequencies of the rotational modes \\
\hline
\multicolumn{2}{c}{Rotation variables} \\
\hline
$l$, $g$, $h$, $L$, $G$, $H$ & Andoyer variables and the moments associated \\
$p$, $r$, $\xi_q$, $P$, $\mathcal{R}$, $\eta_q$ & Modified Andoyer variables and the moments associated \\
\hline
\multicolumn{2}{c}{Rotation outputs} \\
\hline
$\Psi$ & Tidal longitudinal librations \\
$\gamma$ & Physical longitudinal librations \\
$\eta$ & Latitudinal librations \\
$\epsilon$ & Obliquity of Mimas \\
$Q_1+\imath Q_2$ & Polar motion of Mimas \\
\hline
\multicolumn{2}{c}{Tidal data} \\
\hline
$\lambda$ & Dissipative rate \\
$x_i$, $y_i$ & Displacement of the equilibrium \\
\hline
\end{tabular}
\end{table*}

\begin{acknowledgements}
Numerical simulations were made on the local computing ressources (Cluster URBM-SYSDYN) at the University of Namur. This work has been supported by EMERGENCE-UPMC grant (contract number: EME0911).
\end{acknowledgements}

\end{document}